\documentclass[aps,prl,twocolumn,superscriptaddress,showpacs,draft]{revtex4}
\usepackage{graphicx}
\input{epsf}

\newcommand{\rem}[1]{}

\begin{document}

\title{Synchronization and bistability of qubit coupled 
to a driven dissipative oscillator}
\author{O.V.Zhirov}
\affiliation{\mbox{Budker Institute of Nuclear Physics,
630090 Novosibirsk, Russia}}
\author{D.L.Shepelyansky}
\affiliation{\mbox{Laboratoire de Physique Th\'eorique, 
UMR 5152 du CNRS, Universit\'e Toulouse III, 
31062 Toulouse, France}}

\date{October  10, 2007}


\begin{abstract} 
We study numerically the behavior of qubit coupled to a quantum dissipative
driven oscillator (resonator). Above a critical coupling
strength the qubit rotations become synchronized
with the oscillator phase. 
In the synchronized regime, at certain parameters,
the qubit exhibits tunneling between two orientations
with a macroscopic change of number of photons in the resonator.
The life times in these metastable states can be enormously large.
The synchronization leads to a drastic change of qubit radiation
spectrum with appearance of narrow lines 
corresponding to recently observed
single artificial-atom lasing 
[O.~Astafiev {\it et al.} Nature {\bf 449}, 588 (2007)].
\end{abstract}
\maketitle

In physics there are not so many simple quantum problems which are
exactly solvable \cite{morse}. Two of them are monochromatically 
driven two-level atom (spin-half or qubit) \cite{landau} and 
quantum oscillator (unitary or dissipative) \cite{perelomov,weiss}.
One atom weakly coupled to a field in a resonator
is known as the Jaynes-Cummings model which is also integrable
\cite{jaynes,eberly,scully}. At strong coupling 
the dynamics may become nontrivial with the emergence
of classical \cite{zaslavsky,milonni} and quantum chaos
\cite{graham1} but in this case one should have many
atoms which may absorb many photons. For one atom
even with a strong coupling to a quantum photonic field
the dynamics is still relatively simple due to
a total energy balance \cite{graham}.

This old problem of a two-level atom coupled to photons
regained recently a significant interest
due to appearance of long living 
superconducting qubits \cite{vion}
which can be strongly coupled
to a microwave resonator \cite{wallraff,houck,astafiev}.
There are also other possibilities of 
superconducting qubit coupling to a quantum oscillator
\cite{ilichev,buisson}.
The oscillator can be realized as
a  tank circuit tuned to the Rabi frequency \cite{ilichev}
or as a current-biased  dc SQUID \cite{buisson}
allowing efficient energy exchange with a qubit.
Possibilities of qubit coupling to a cooled nanomechanical resonator
are actively discussed \cite{cleland,rugar}
and coupling between micro-mechanical cantilever and 
atomic spin found impressive experimental implementations
(see \cite{berman} and Refs. therein).
However, the most intriguing way seems to be 
the coupling with a microwave resonator
where a lasing has been realized recently 
with 6 - 30 photons pumped into the resonator \cite{astafiev}.
Contrary to recent interesting theoretical studies
\cite{berman,rodrigues,shnirman} where pumping is applied to a 
spin or qubit
we concentrate here on the case where a monochromatic
pumping is applied to a  dissipative oscillator (resonator).
Such a dissipative oscillator can be also viewed as
a semiclassical detector which performs monitoring of a qubit.
This continuous type of  measurements is now actively discussed  
for superconducting qubits and other 
solid-state devices \cite{korotkov}.
The continuous measurement of a superconducting qubit
is realized in \cite{ilichev}.

\begin{figure}
\epsfxsize=8.5cm
\epsffile{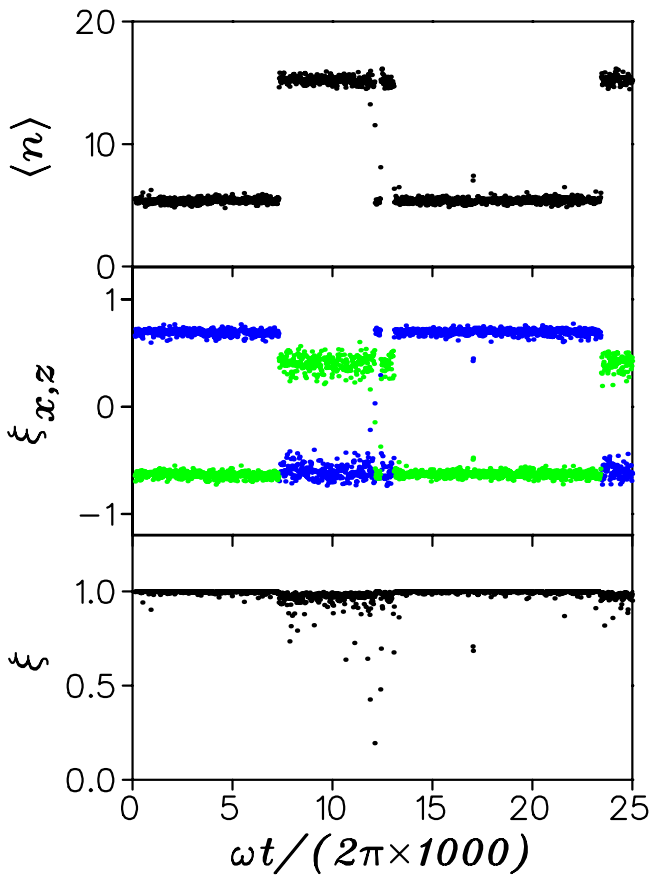}
\caption{(color online) Bistability of qubit coupled to a driven oscillator
with jumps between two metastable states. Top panel shows 
average oscillator level number $\langle n \rangle$ as a function of time $t$
at stroboscopic integer values $\omega t/2\pi$;
middle panel shows the qubit polarization
vector components $\xi_x$ (blue/black) and $\xi_z$ (green/gray)
at the same moments of time; the bottom panel shows 
the degree of qubit polarization $\xi$. 
Here the system parameters are $\lambda/\omega_0=0.02$,
$\omega/\omega_0=1.01$, $\Omega/\omega_0=1.2$, 
$f=\hbar \lambda \sqrt{n_p}$, $n_p=20$ and $g=0.04$.
}
\label{fig1}
\end{figure}

The Hamiltonian of our model reads
\begin{equation}
  \hat{H} = \hbar \omega_0  \hat{n}  
  - \hbar \Omega \sigma_x /2
  + g \hbar \omega_0 ( \hat{a} + \hat{a}^{\dag}) \sigma_z 
  + f \cos{\omega t} \left( \hat{a} + \hat{a}^{\dag} \right) 
\label{eq1} 
\end{equation}
where $g$ is a dimensionless coupling constant,
the driving force amplitude and frequency are 
$f=\hbar \lambda \sqrt{n_p}$ and  $\omega$,
the oscillator frequency is $\omega_0$ and $\hbar \Omega$
is the qubit energy spacing. As in \cite{shnirman}
we choose a qubit coupling via $\sigma_z$. We assume that
the qubit life time is enormously long and
that its dynamics is perturbed only by the coupling with
the driven dissipative oscillator. The dissipation rate of oscillator
is $\lambda$ and we assume the quality factor to be 
$Q=\omega_0/\lambda \sim 100$. The evolution of the whole system
is described by the master equation for the density matrix $\hat{\rho}$ 
which has the standard form \cite{weiss}:
\begin{equation}
\dot{\hat{\rho}} = -  \frac{i}{\hbar} [\hat{H},\hat{\rho}] +
\lambda ( \hat{a}  \hat{\rho} \hat{a}^{\dag} 
- \frac{1}{2} \hat{a}^{\dag}  \hat{\rho} \hat{a} 
- \frac{1}{2} \hat{\rho} \hat{a}^{\dag}  \hat{a})
\label{eq2} 
\end{equation}
The numerical simulations are done by direct solution of time
evolution of $\hat{\rho}$ expanded in a finite basis of oscillator
states $n$, by the state diffusion method  \cite{stdif} 
and by the method of Quantum Trajectories (QT) \cite{percival}.
We ensured that these methods give the same results
but the majority of data is obtained with quantum trajectories
which we found to be more suitable for massive simulations.
In addition the QT have an advantage of providing 
a pictorial illustration of individual
experimental runs. The numerical details 
are the same as in \cite{benenti,zhirov} and
we use here up to $n=70$ oscillator states which give
good numerical convergence. Our results show that 
a coupling of two simple integrable models
gives a nontrivial interesting behavior.

A typical example of QT is shown in Fig.~\ref{fig1}.
It shows two main properties of the evolution:
the oscillator spends a very long time at some average level 
$\langle n \rangle = n_-$
and then jumps to another significantly different  value $n_+$.
At the same time  the polarization vector of qubit $\vec{\xi}$
defined as $\vec{\xi} = Tr(\hat{\rho} \vec{\sigma})$
also changes its orientation direction
with a clear change of sign of $\xi_x$ from $\xi_x >0$ to
$\xi_x<0$.
The time averaged values of $\xi_{y,z}$
are zero but when they are taken at stroboscopic 
integer moments $\omega t/2\pi$
they also show transitions between two metastable states.
The transition time is approximately $t_m \sim 1/\lambda$
being rather small compare to the life time
in a metastable state. Inside such a state the degree
of qubit polarization $\xi=| \vec{\xi} | $
is very close to unity showing that the qubit remains mainly in a pure state.
The drops of $\xi$ appear only during transitions
between metastable states. Special checks show that an
inversion of $\xi_x$ by an additional pulse (e.g. from $\xi_x >0$
to $\xi_x <0$)
produces a transition of oscillator to a corresponding
state (from $n_-$ to $n_+$) after time $t_m  \sim 1/ \lambda$.
Thus we have here an interesting situation when a quantum flip of qubit
produces a marcoscopic change of a state of detector (oscillator) 
which is continuously coupled to a qubit
(we checked that even larger variation $n_{\pm} \sim n_p$ is possible
by taking $n_p=40$). In addition to that inside a metastable
state the coupling induces a {\it synchronization}
of qubit rotation phase with the oscillator phase
which in its turn is fixed by the phase of driving field.
The synchronization  is a universal phenomenon for
classical dissipative systems \cite{pikovsky}.
It is known that it also exists for dissipative quantum systems
at small effective values of $\hbar$ \cite{zhirov}. 
However, here we have a new unusual case of qubit synchronization
when a semiclassical system produces synchronization
of a pure quantum two-level system.

\begin{figure}
\centerline{\epsfxsize=7.5cm\epsffile{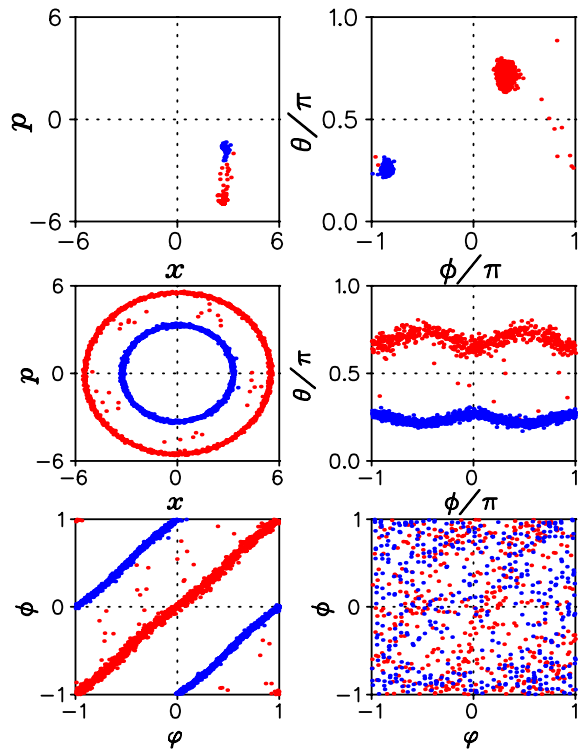}}
\vglue -0.0cm
\caption{(color online) Top panels:  the Poincar\'e section
taken at integer values of $\omega t/2\pi$
for oscillator with 
$x=\langle (\hat{a}+\hat{a}^{\dag})/\sqrt{2} \rangle$,
$p=\langle (\hat{a}-\hat{a}^{\dag})/\sqrt{2}i \rangle$ (left) and 
for qubit polarization 
with polarization angles $(\theta,\phi)$ defined in text (right).
Middle panels: the same quantities shown at irrational moments
of $\omega t/2\pi$.
Bottom panels: the qubit polarization phase $\phi$
{\it vs.} oscillator phase $\varphi$
($p/x= - \tan \varphi$)
at time moments as in  middle panels
for $g=0.04$ (left) and $g=0.004$ (right). 
Other parameters and the time interval are as in Fig.~\ref{fig1}.
The color of points is blue/black for $\xi_x >0$
and red/gray for $\xi_x <0$. 
}
\label{fig2}
\end{figure}

The phenomenon of qubit synchronization is illustrated in a more clear
way in Fig.~\ref{fig2}.  The top panels 
taken at integer values $\omega t/2\pi$ 
show the existence of two fixed points
in the phase space of oscillator (left) and qubit (right)
coupled by quantum tunneling
(the angles are determined as 
$\xi_x=\xi \cos \theta, \xi_y=\xi \sin \theta \sin \phi,  
 \xi_z = \xi \sin \theta \cos \phi$).
A certain scattering of points in a spot of finite size
should be attributed to quantum fluctuations.
But the fact that on enormously long time (Fig.~\ref{fig1})
the spot size remains finite clearly implies
that the oscillator phase $\varphi$ is locked with the driving
phase $\omega t$ inducing the qubit synchronization 
with  $\varphi$ and $\omega t$. The plot at $t$ values incommensurate with
$2\pi/\omega$ (middle panels) shows that in time the oscillator
performs circle rotations in $(p,x)$ plane with frequency $\omega$
while qubit polarization rotates around $x$-axis with the
same frequency. Quantum tunneling gives transitions between
two metastable states.  The synchronization of 
qubit phase $\phi$ with oscillator phase $\varphi$
is clearly seen in bottom left panel  where points form 
two lines corresponding to two metastable states.
This synchronization disappears below a certain critical
coupling $g_c$ where the points become scattered over the whole plane
(panel bottom right). It is clear that quantum fluctuations
destroy synchronization for $g < g_c$.
Our data give $g_c  \simeq 0.008$ for
parameters of Fig.~\ref{fig1}.

\begin{figure}
\centerline{\epsfxsize=4.2cm\epsffile{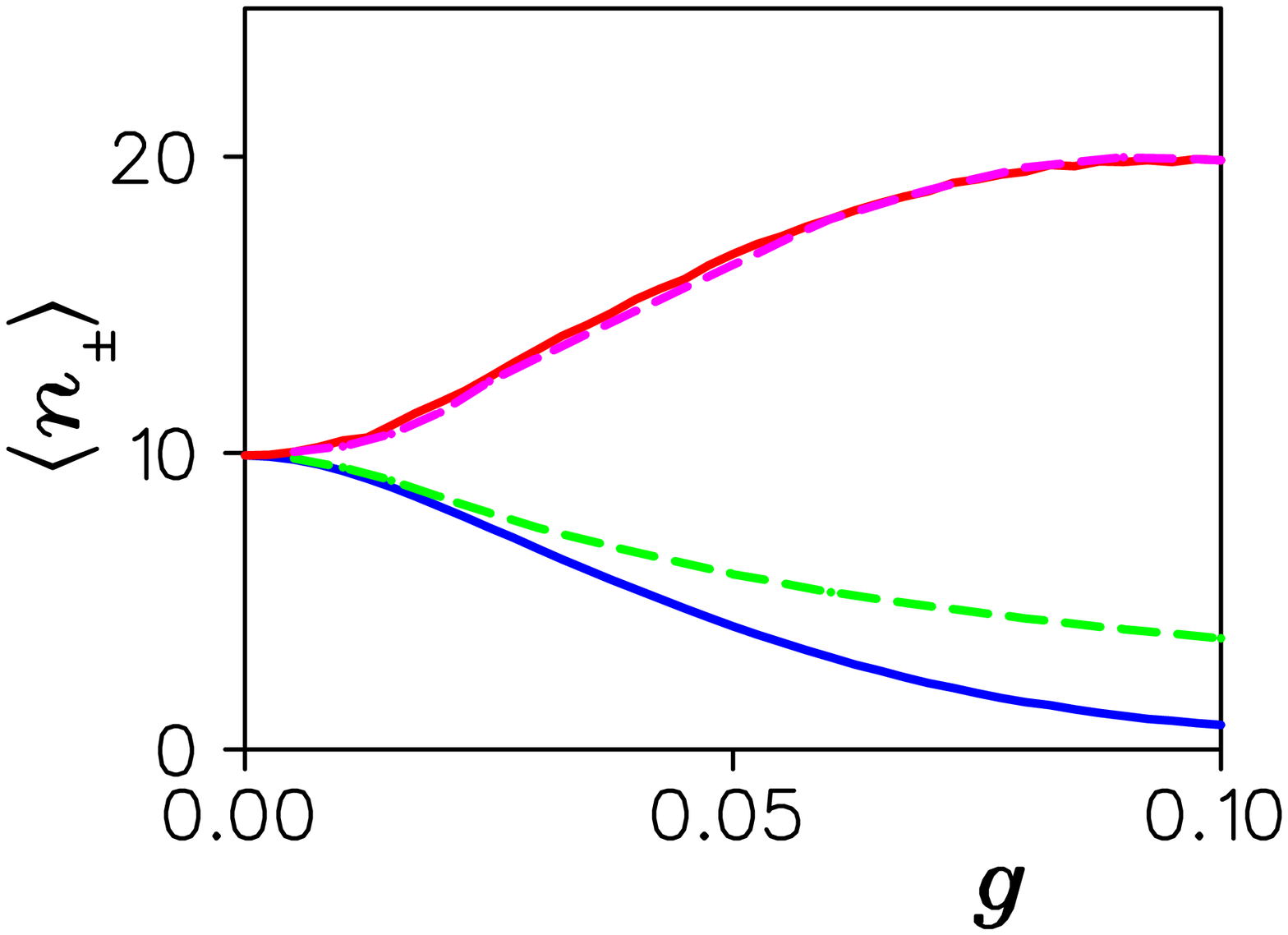}
\hfill\epsfxsize=4.2cm\epsffile{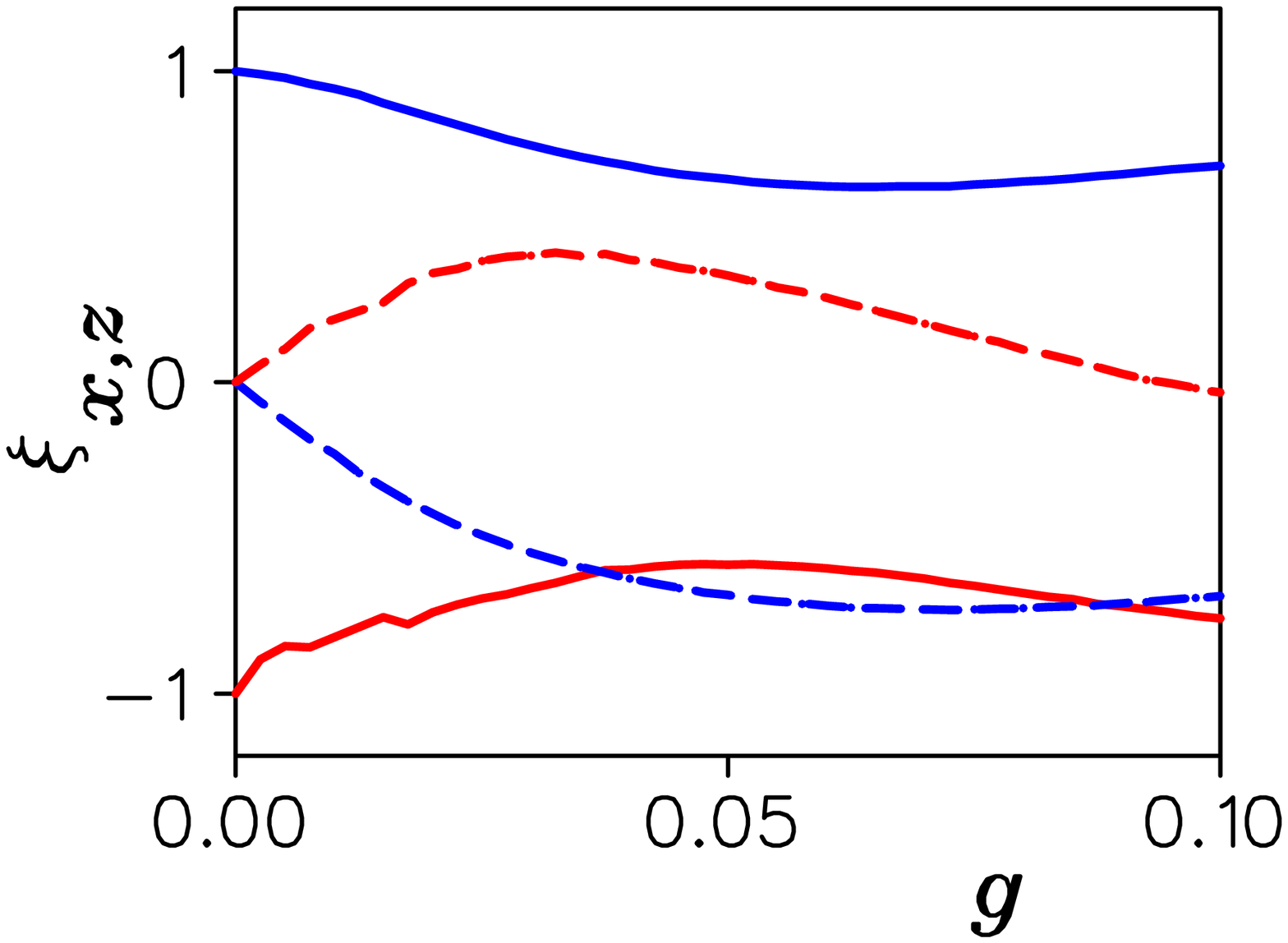}}
\vglue -0.0cm
\caption{(color online) Right panel:
dependence of average qubit polarization components 
$\xi_x$ and $\xi_z$ (full and dashed curves) on $g$,
averaging is done over stroboscopic times (see Fig.\ref{fig1}) in the interval
$100 \leq \omega t/2\pi \leq 2 \times 10^4$;
color  is fixed by the sign of $\xi_x$
averaged over 10 periods 
(red/gray for  $\xi_x<0$, blue/black for  $\xi_x>0$;
this choice fixes also the color on right panel).
Left panel: dependence of 
average level of oscillator  in two metastable states
on coupling strength $g$,
the color is fixed by the sign of $\xi_x$ on right panel
that  gives red/gray for large $n_+$ and blue/black for  small $n_-$;
average is done over the quantum state and stroboscopic times
as in the left panel;
dashed curves show theory dependence
(see text)). Two QT are used with initial value $\xi_x=\pm1$.
All parameters are as in Fig.~\ref{fig1}
except $g$.
}
\label{fig3}
\end{figure}

\begin{figure}
\centerline{\epsfxsize=7.5cm\epsffile{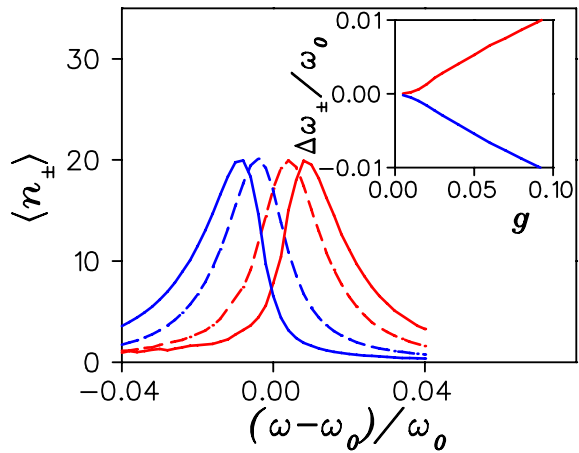}}
\vglue -0.0cm
\caption{(color online) 
Dependence of average level $n_\pm$  of oscillator in two metastable states
on the driving frequency $\omega$
(average and color choice are the same as in right panel of  Fig.~\ref{fig3});
coupling is $g=0.04$ and $g=0.08$ (dashed and full curves).
Inset shows the variation of  position of maximum at $\omega=\omega_\pm$
with coupling strength $g$, $\Delta \omega_\pm = \omega_\pm - \omega_0$.
Other parameters are as in Fig.~\ref{fig1}.
}
\label{fig4}
\end{figure}

The variation of bistable states with coupling strength $g$
is shown in Fig.~\ref{fig3}. The difference between $n_+$ and $n_-$
grows with $g$. It is striking that $n_-$ may become close to zero.
The direction of qubit polarization also changes in a 
smooth but nontrivial way.
It is also important to note that according to our data
the dispersion of oscillator wave function in metastable states 
is compatible with the dispersion of a coherent state
with $n_\pm$.
This corresponds to a wave packet collapse induced by
dissipation (see \cite{benenti,zhirov} and Refs. therein).
The dependence 
of $n_{\pm}$ on driving frequency $\omega$ 
is shown in Fig.~\ref{fig4}. A symmetric double peak structure
is evident: for $\omega > \omega_0$ the metastable state with 
$\xi_x <0$ has maximal $n$ value while for
$\omega < \omega_0$ the state with maximal $n$
has $\xi_x >0$ (note color interchange). The peak width
is approximately equal to the dissipation rate $\lambda$.
With the increase of $g$ their form becomes asymmetric
indicating importance of nonlinear effects.
The splitting of peaks grows approximately linearly with $g$
(inset at Fig.~\ref{fig4}) and
reminds  the vacuum Rabi splitting effect \cite{eberly}.
The shift $\Delta \omega_{\pm} $
explains two states $n_\pm$ of driven oscillator
well described by 
$n_\pm = n_p \lambda^2/(4(\omega - \omega_0 -\Delta \omega_\pm)^2+\lambda^2)]$ 
(see dashed curves in Fig.~\ref{fig3} left  traced
with numerical values of $\Delta \omega_\pm $ from Fig.~\ref{fig4} inset).
To estimate $\Delta \omega_\pm$ we note that the frequency of 
effective Rabi oscillations between quasi-degenerate
levels is $\Omega_R \approx g \omega_0 \sqrt{n_\pm+1}$ \cite{jaynes,scully}
that gives $\Delta \omega_\pm  \approx d \Omega_R/dn 
\approx \pm g \omega_0/2\sqrt{n_\pm+1}$
in a good agreement with data of Fig.~\ref{fig4} for moderate $g$.

\begin{figure}
\centerline{\epsfxsize=7.0cm\epsffile{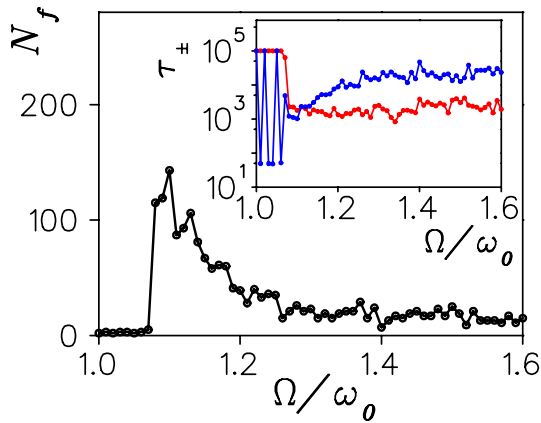}}
\vglue -0.0cm
\caption{(color online) 
Dependence of number of transitions $N_f$ between metastable states
on rescaled qubit frequency $\Omega/\omega_0$ 
for parameters of Fig.~\ref{fig1}; $N_f$ are computed along 2 QT of
length $10^5$ driving periods. Inset shows life time
dependence on $\Omega/\omega_0$ for two metastable states
($\tau_+$ for red/gray, $\tau_-$ for blue/black,
 $\tau_\pm$ are given in number of driving periods; color choice is as 
in Figs.~\ref{fig2},\ref{fig3}).
}
\label{fig5}
\end{figure}

The properties of two metastable states are analyzed in Fig.~\ref{fig5}.
The number of transitions $N_f$ between these states 
has a pronounced peak at $\Omega \approx 1.1 \omega_0$
that approximately corresponds to a resonance condition
$\Omega - \omega \approx 2g\omega_0$. 
For $\Omega < 1.08 \omega_0$
there is an abrupt drop of $N_f$ and bistability 
becomes irregular disappearing for certain $\Omega$,
but the synchronization still remains.
Quite interestingly, the data show that for
$\Omega > 1.1 \omega_0$ the life times of each state
are rather different and enormously large, 
generally $\tau_- > \tau_+ \gg \omega_0/\lambda$.

\begin{figure}
\centerline{\epsfxsize=4.2cm\epsffile{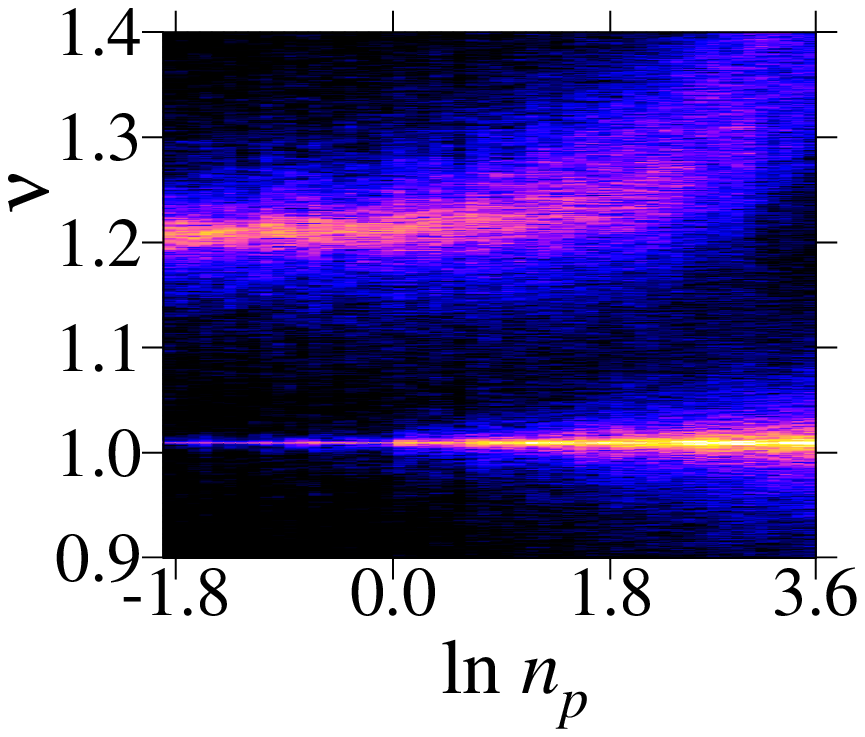}
\hfill\epsfxsize=4.2cm\epsffile{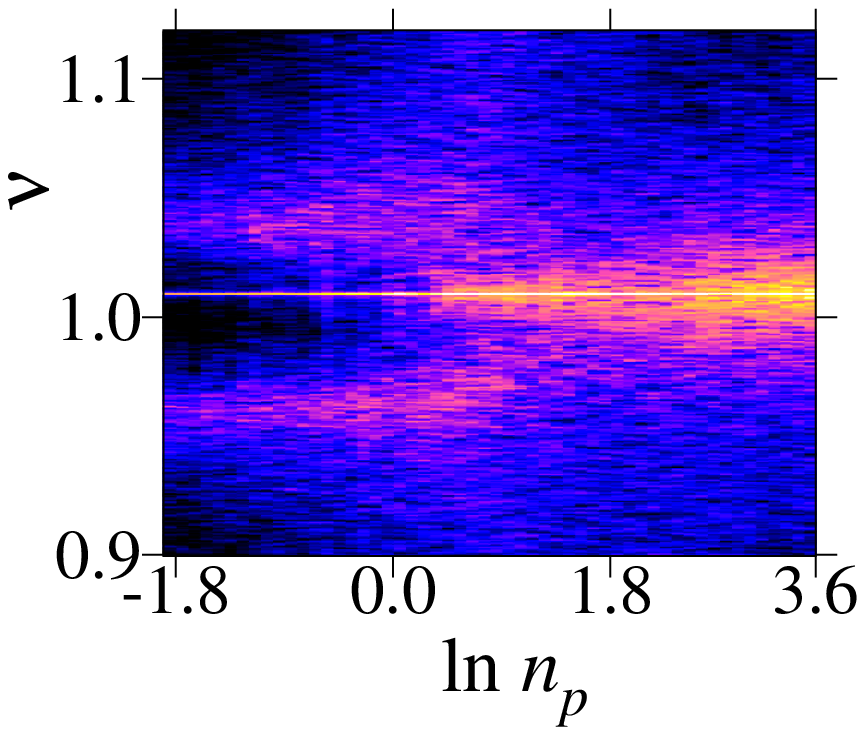}}
\vglue -0.0cm
\caption{(color online) Spectral density $S(\nu)$
of qubit radiation $\xi_z(t)$ as function of driving power $n_p$
in presence of phase noise in $\phi$ with diffusion rate 
$\eta = 0.004 \omega_0$.
Left: $\Omega/\omega_0=1.2$;
right:  $\Omega/\omega_0=1$.
Other parameters are as in Fig.~\ref{fig1}.
Color/grayness shows $S(\nu)$
in logarithmic scale 
(white/black for maximal/zero), $\nu$ is given in units of $\omega_0$.
}
\label{fig6}
\end{figure}

The spectrum of qubit radiation $S(\nu)$ 
in presence of phase noise in $\phi$ is shown in Fig.~\ref{fig6}.
It confirms the main features discussed above: for $\Omega/\omega_0=1.2$
the growth of driving power $n_p$ induces the synchronization
of qubit with radiation suppression at qubit frequency $\Omega =1.2 \omega_0$
and appearance of narrow line with
lasing at $\nu=\omega$.
For $\Omega =\omega_0$   the radiation spectrum $S(\nu)$ 
at $n_p<1$ has two broad peaks at $\nu=\omega \pm g \omega_0$
corresponding to the
vacuum Rabi splitting  \cite{eberly}
(narrow line from driving source at $\nu=\omega$ 
is also visible in this case).
At strong driving $n_p > 1$ the synchronization takes place
with appearance of one lasing line at $\nu = \omega$.
For both values of $\Omega$ the transition to 
synchronization/lasing takes place at $n_p > n_{pl} \approx 2$.
The spectrum $S(\nu)$ in Fig.~\ref{fig6} has close 
similarities with the  spectrum
observed recently in a single artificial-atom lasing
\cite{astafiev} which appears at 
a similar threshold $n_{pl} \approx 1$.
A shift related to  splitting $\omega_\pm \approx g \omega_0/2\sqrt{n}$
is also seen experimentally.
However, the vacuum Rabi splitting is not visible in
Fig.3c of \cite{astafiev}. 
Exact comparison requires much more extended numerical simulations
since in   \cite{astafiev} $Q \approx 10^4$ while we have $Q \sim 100$. 

In conclusion our numerical studies show a nontrivial
behavior of a rather simple model given by Eqs.~(\ref{eq1},\ref{eq2}).
It is characterized by bistability and synchronization of
qubit induced by its coupling to a quantum driven dissipative
oscillator. As for the vacuum Rabi splitting \cite{eberly}
it is important that the oscillator is quantum since 
the effect is absent for
a classical dissipative oscillator with commuting $a, a^{\dag}$ (\ref{eq1}). 
A better analytical description and understanding
of the behavior discussed requires further studies.
Especially interesting is the analysis of long life times $\tau_\pm$ 
in metastable states related to a macroscopic quantum tunneling.

Preparing the paper to submission we became aware
of the preprint \cite{girvin} where a similar model is studied
but synchronization is not discussed there.

This work is supported by the EC  project EuroSQIP
and RAS joint scientific program  "Fundamental problems of nonlinear dynamics"
(for OVZ).

\end{document}